\documentclass[letter,conference]{IEEEtran}

\usepackage[pdftex]{graphicx}
\graphicspath{{../graphics/}}
\DeclareGraphicsExtensions{.pdf,.jpeg,.png}
\usepackage{subfigure}
\usepackage{cite}
\usepackage[colorlinks,linkcolor=black,citecolor=black,urlcolor=black]{hyperref}    % Creates hyperlinks from ref/cite 
\usepackage{url}
\usepackage{textcomp}           % More symbols
\usepackage{color}
\usepackage{threeparttable}
\usepackage{float}
\usepackage{booktabs}

\usepackage{pifont}
\makeatletter
\renewenvironment{description}
               {\list{}{\labelwidth\z@ \itemindent-\leftmargin
                        }}
               {\endlist}
\renewcommand*\descriptionlabel[1]{\hspace\labelsep
                                 \normalfont\bfseries #1}
\makeatother

\makeatletter
\def\paragraph{\@startsection{paragraph}{4}{1\parindent}{0ex}{0ex}{\normalfont\normalsize\itshape}}%
\makeatother

\title{The Role of the Internet of Things\\ in Network Resilience}
\author{
\IEEEauthorblockN{Hauke Petersen\IEEEauthorrefmark{1},
Emmanuel Baccelli\IEEEauthorrefmark{2},
Matthias W\"ahlisch\IEEEauthorrefmark{1},
Thomas C. Schmidt\IEEEauthorrefmark{3},
Jochen Schiller\IEEEauthorrefmark{1}
}
\IEEEauthorblockA{Freie Universit\"at Berlin, Germany\IEEEauthorrefmark{1}
\quad
INRIA, France\IEEEauthorrefmark{2}
\quad
HAW Hamburg, Germany\IEEEauthorrefmark{3}
}
\IEEEauthorblockA{\{first.last\}@fu-berlin.de,
emmanuel.baccelli@inria.fr,
t.schmidt@ieee.org}
}

\begin{document}
% make the title area
\maketitle

\begin{abstract}
Disasters lead to devastating structural damage not only to buildings and transport infrastructure, but also to other critical infrastructure, such as the power grid and communication backbones. Following such an event, the availability of minimal communication services is however crucial to allow efficient and coordinated disaster response, to enable timely public information, or to provide individuals in need with a default mechanism to post emergency messages. The Internet of Things consists in the massive deployment of heterogeneous devices, most of which battery-powered, and interconnected via wireless network interfaces. Typical IoT communication architectures enables such IoT devices to not only connect to the communication backbone (i.e. the Internet) using an infrastructure-based wireless network paradigm, but also to communicate with one another autonomously, without the help of any infrastructure, using a spontaneous wireless network paradigm. In this paper, we argue that the vast deployment of IoT-enabled devices could bring benefits in terms of data network resilience in face of disaster. Leveraging their spontaneous wireless networking capabilities, IoT devices could enable minimal communication services (e.g. emergency micro-message delivery) while the conventional communication infrastructure is out of service. We identify the main challenges that must be addressed in order to realize this potential in practice. These challenges concern various technical aspects, including physical connectivity requirements, network protocol stack enhancements, data traffic prioritization schemes, as well as social and political aspects.
\end{abstract}


%\IEEEpeerreviewmaketitle

\section{Introduction}
\label{sec:introduction}

Every year witnesses large-scale disasters around the world, affecting millions of people. A crucial aspect of crisis management is distribution of information, immediately after the disaster occurs. Usually, we rely on data communication networks to deliver information fast, reliably, anywhere, anytime. The Internet is today's communication backbone, used not only for transferring data but it is also utilized as back-end for voice communication \cite{baldwin2010}. Even though the Internet is a highly interconnected system with several backup paths, it is vulnerable to the effects of large scale disasters, which can lead to local but also global communication outages and thus significant disruption of crisis management after such disaster occurs.

%Most of these disasters are caused by natural forces such as storms, floods, and earthquakes, but they are also triggered by artificial threats such as power-grid blackouts. As for most of the cases we do not have any possibility to prevent disasters from happening, leaving us only with the choice to mitigate the effects. For this an efficient and coordinated disaster response relies heavily on the availability of communication services.

In large scale disaster scenarios, typical approaches to (re)establish communication abilities yield manual installation of new hardware, which takes time. However, massive deployment of heterogeneous, Internet-enabled embedded devices is taking place, amounting to what is called the Internet of Things (IoT) \cite{atzori2010}. A large part of these devices is battery powered and communicate wirelessly. Predictions show that their number will grow reach billions over the next decade \cite{ericcson2011, vermesan2011}, and will result in a very dense deployment which will significantly reshape the Internet's edge architecture, allowing for more decentralized and dynamic communication paradigms.

In this paper, we discuss to which extent the Internet of Things may increase network resilience in disaster scenarios. We argue that stakeholders---in particular the general public--- would significantly benefit from leveraging the decentralized nature of the Internet of Things, that could enable minimal communication services in scenarios where the conventional communication infrastructure is out of service. We analyze the main challenges that must be addressed in order to realize this potential. These challenges concern various technical aspects, including physical connectivity requirements, network protocol stack enhancements, data traffic prioritization schemes, as well as social and political aspects, that we detail in the following.

\section{Current Communication in Disaster Scenarios}
\label{sec:disasterrequirements}

Communication in disaster scenarios is primarily driven by exchanging \emph{important} instead of arbitrary information. Different groups of actors have different communication requirements, which finally lead to the deployment of the underlying technology.

%packet based data network implemented in the Internet \cite{baldwin2010}. 
%Additionally to the Internet there exists a number of separated networks as SIRPNet, CHRONOS or JWICS. But as these are used for a limited audience or are even substituted by secured channels over the Internet, we will use the Internet as synonym for communication networks in general in this article.

\subsection{Communication requirements}

A disaster may disconnect a complete country from the rest of the world or limit capacities to data with very low throughput. Ideally this remaining connectivity should be used by the most important services and actors, mainly for information-sharing and coordination. With passing time after a disaster happened these priorities are further subject to change. In the period of time following the initial impact the actual saving of human life is the most important action that needs to be coordinated. This is generally done between first responders such as fire-fighters, police, and technical response forces. For disaster with devastating impacts the prioritization of communication capabilities will shift after the initial time period towards governmental organizations and non-governmental organizations that are concerned with providing foot and shelter and restoring the social systems.

During all phases there is further need of communication for the general public. The population in the affected areas has to be warned of threats and to be informed of retreat routes and similar information. People need further to communicate with relatives and other persons inside as well as outside the disaster area to check on their status \cite{pl-cccaf-07}.

The actors that operate in disaster areas and their used communication mechanisms can be categorized as follow.

\begin{description}
\item[First responders:]

Communication between teams using voice and text connections.

\item[Governmental organizations:]

Communication between central situation centers using voice, text and further access to databases.

\item[Non-governmental organizations:]

Coordination using voice and text communication, access to logistical databases.

\item[The press:] Sending texts out of the disaster area, audio and video broadcasts.

\item[The general public:]
Emergency calls, status calls, receive news and situational updates, receive environmental/emergency warnings.
\end{description}

\begin{figure}
  \centering
  \includegraphics[width=\columnwidth]{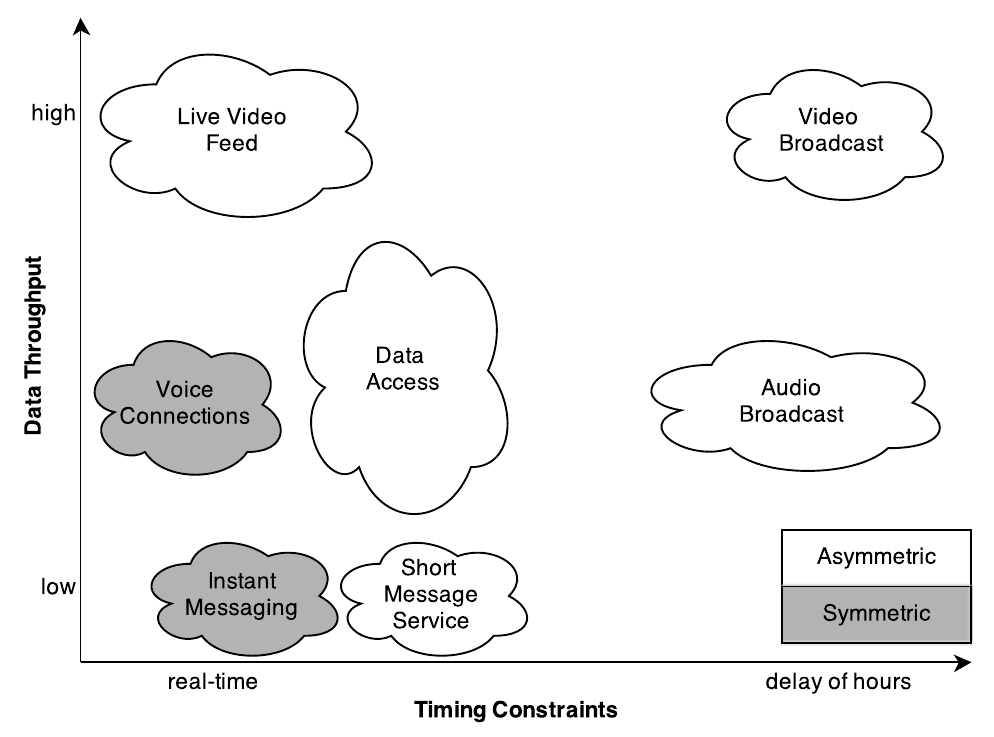}
  \caption{Comparison of delay and throughput requirements of typical applications in disaster scenarios.}
  \label{fig:matrix}
\end{figure}

We compare the communication services in a three dimensional space with respect to basic communication parameters, needed throughput, the direction in which the data flows, and finally the requirements on timing constraints (cf., Figure~\ref{fig:matrix}).

It is worth noting that the distinction between the actors is not exclusive. In particular, the general public covers multiple fields. With the advent of blogging, social networks, and micro messaging (e.g., Twitter) \emph{citizen journalism} has been established to complement the press by public contributions. After the Tohoku earthquake in 2011, for example, $\approx$ 50~\% of the photos related to the disaster in the Tokyo area have been uploaded to Flickr in less than 24 hours. This information fulfills two purposes, it informs other people about the current state but also helps rescue teams to identify relevant areas. Previous disasters also shown that first responders are not only experts but also volunteers from the neighborhood, who help~\cite{pl-cccaf-07}. These observations have direct implications on the devices, which are used on-site, and thus on the deployed technology. Professionals such as press, NGOs, and first responders may own special hardware. The general public is equipped with mass market devices (e.g., smartphones) providing basic communication functionality. Building a more robust communication infrastructure should consider this and incorporate public devices.

The distance between two communication partners, which needs to be bridged, is diverse even within a group of stakeholders. Typical NGO scenarios illustrate this nicely. Field workers require short range communication between peers, as well as long range communication to request external data and to interact with external operation control center. Short range communication is limited to a smaller geographic area, in which long range communication bridges further distances. The latter is currently implemented in the Internet.

%=> hieraus device requirements etc. ableiten: typical first responders come with special equipment, general public has smartphone

\begin{figure*}
  \centering
  \subfigure[Text-based communication]{\includegraphics[width=\columnwidth]{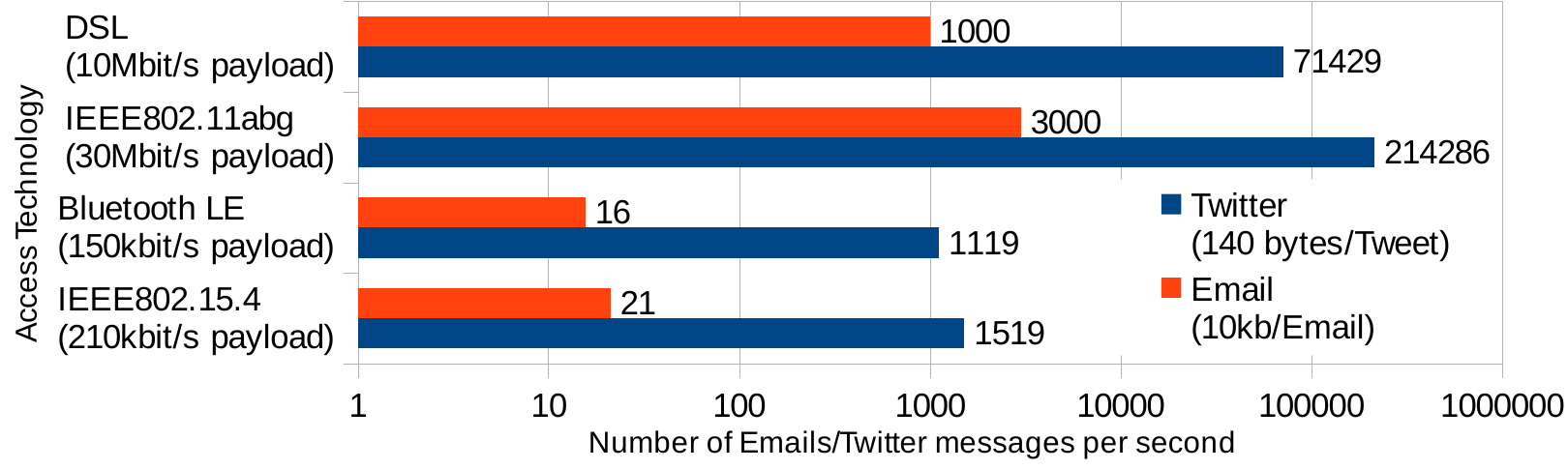}}
  \quad
  \subfigure[Voice and video calls]{\includegraphics[width=\columnwidth]{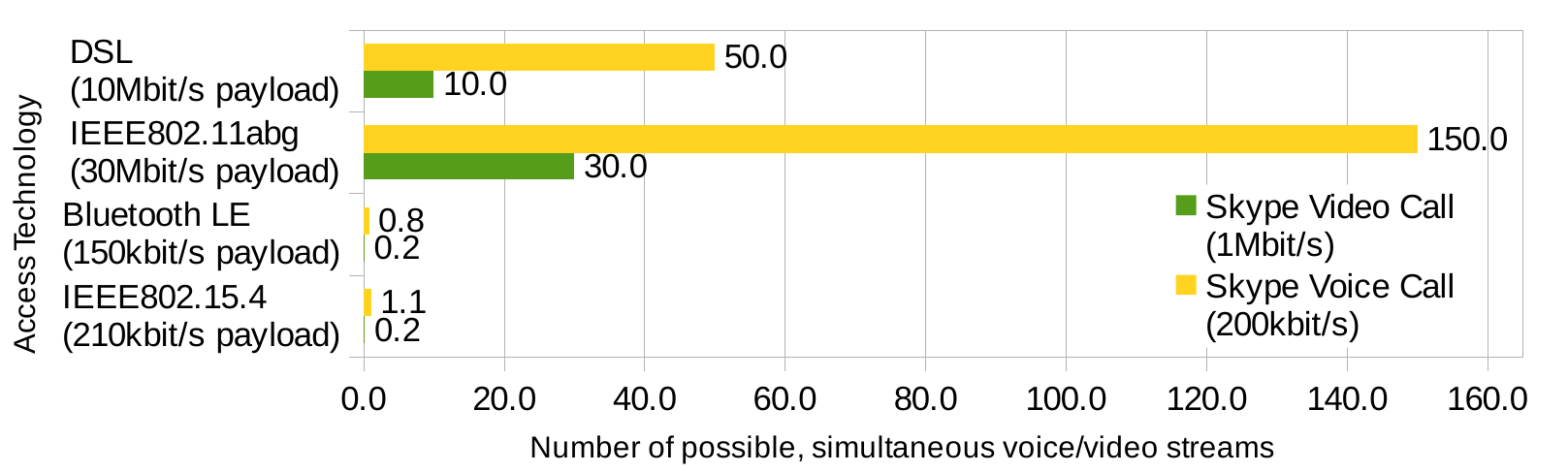}}
  \caption{Usage of available throughput between two directly connected peers by typical applications for different access technologies.}
  \label{fig:cmpthroughput}
\end{figure*}

\subsection{Dependency on fixed infrastructure}
\label{sec:dependencyonfixedinfrastructure}
Today's communication is heavily based on the Internet. Originally, different infrastructures have been operated for voice and data traffic. This distinction continuously converges towards a unified backbone implemented by the Internet \cite{baldwin2010}. The Internet provides packet-based data delivery and allows for a wide range of communication services on top of the delivery infrastructure, making it more attractive compared to other backbones.

Successful communication in disaster scenarios is tied to the successful operation of the Internet. This relates to two perspectives, the outsider and the insider perspective. A disaster that affects Internet infrastructure components may also affect people living in areas which are geographically outside of the disaster region. During the localized 9/11 attack smaller Internet outages have been experienced in Japan, for example. Given that the Internet is the backbone of our daily communication this can lead to severe problems. People inside the disaster area rely on the Internet (or Internet technologies) to exchange information.

The proper operation of the basic Internet infrastructure depends on wired connections and fixed power supplies. Both components make the Internet vulnerable to breakdown caused by disasters. Large scale disasters by definition have in common that wide areas of land are affected by immense forces such as floods, storms, or earthquakes. These forces lead usually to an immense destruction of man-made infrastructure, which is also important for the Internet backbone. Buildings accommodating points of presence collapse (e.g., 9/11), oversea cables break (e.g., Japan earthquake 2011), or power supplies turn down (e.g., Italy blackout 2003), for example. Satellite Internet access replaces cables in specific regions but those equipment still represent rather fix component.

For Internet hardware, it can be distinguished between two basic classes of fault modes that leave the infrastructure in a non-working state: Systems can suffer recoverable fault or they can suffer permanent damage. For the first class typical fault modes are power outages and overload conditions. As soon as power is restored or overload conditions are resolved, the system can continue in normal operation and little intervention by the network operators is required. Typical types for the second class of fault modes are broken wires and physically damaged hardware. In both cases massive repair effort by technical personal is required, as hardware needs to be replaced or connections have to be rewired. In case of highly destructive disasters these fault modes are more common. During most disasters buildings and power grids collapse, and the repairs requires significant time. Both fault classes are in fact not independent of each other. The outage of a backbone router will lead to a redirection of traffic which can lead to an overload condition and subsequent failure of another router.

The 2003 Italy blackout demonstrated the consequences of long-range, cascading failures and the interplay between the Internet and the fix power grid. A storm caused cascading outage of several power stations, which caused a failure of the Internet infrastructure, finally leading to additional breakdowns of power stations. 

A fast recovery of communication infrastructure is of utmost importance. The common approaches today are to set-up temporary connectivity using mobile 3G/GSM base stations, satellite up-links, and improvised wiring paired with mobile generators for power supply. All these techniques though have in common that considerable time is needed to set them up. Depending on the location of the disaster, the (heavy) equipment needs to be transported, deployed, and initialized. For the time this takes the connectivity in the disaster area is very limited with respect to reachability and capacity. Furthermore, in the meantime privately installed wireless infrastructure may conflict with regained communication networks. The Haiti earthquake 2010 strikingly illustrated this when local ISPs restored 90~\% of the network using wireless technology but Non-Governmental Organizations (NGOs) accidentally broke network communication by taking over the wireless spectrum. 

\subsection{Towards a disaster-adaptive communication infrastructure}

Without doubt the Internet is very fundamental to enable communication---before, during, and after a disaster happened. Even though the Internet is a highly connected infrastructure providing high redundancy, its resilience is currently limited due to very basic dependency on fixed infrastructure components. Evolving the Internet to a completely disaster agnostic infrastructure with full service capabilities is a rather unrealistic challenge even when applying future Internet technologies. However, narrowing the scope to \emph{minimal communication} reduces complexity and complies with the principle needs in disaster scenarios.

To overcome the major dependency on fixed components, communication networks are complemented by wireless transmission and battery power. The Internet of Things (IoT) inherently implements this perspective. On the downside, wireless technology and low energy result in constrained throughput. For typical IoT access technologies Figure~\ref{fig:cmpthroughput} clearly indicates that still a reasonable amount of messages and calls can be exchanged between two parties. Building a disaster resilient communication network which provides these communication abilities in a stable deployment but with the flexibility of the Internet improves the current state of art.

%make clear, we don't need the full service flexibility of the Internet, first step basic communication

%people: joint decision, small messages are enough (see CHI)

%After a disaster typically we have two scenarios (a) NGOs with special equipment, (b) use daily equipment

\section{Resilience Potential of the Internet of Things}
\label{sec:iotpotential}

The number of devices connected to the Internet has seen a steady growth since its creation. In the 90s, this growth was fueled by the advent of the hypertext transport protocol and the web. In the 2000s, this growth was driven by the new availability of wired broadband Internet access which enabled other popular applications such as multimedia streaming. Over the last decade, the growth has been driven by the emergence of wireless broadband Internet access via cellphones, laptops, tablets, and by novel, ultra-connected applications such as social networks. It is now projected that the growth will be fueled by the Internet of Things (IoT), i.e. the massive deployment of heterogeneous, communicating devices \cite{vermesan2011, ericcson2011}, ranging from wireless sensors to smart home appliances, which will blend in the global network, challenging the traditional notions of 'Internet host' and 'router'.

A significant part of the IoT thus consists in billions of battery powered devices that can communicate wirelessly, deployed in every location where humans shape their environment. In fact, most IoT devices use a communication architecture that is fundamentally richer that the conventional, infrastructure-based communication architecture employed to date. By leveraging a spontaneous wireless networking paradigm \cite{ccb-sigcomm-13}, such IoT devices are natively able to both (i) communicate via access points of the infrastructure if they are available, and (ii) communicate with one another autonomously, without the infrastructure as intermediary, if the latter is not available. Spontaneous wireless networking provides the necessary automatic mechanisms so that IoT devices can dynamically self-organize the relaying of data towards destination \cite{sundmaeker2010}. In that sense, each such IoT device is by default both host and router.

Thus, when one considers the IoT as a dense collection of battery-powered devices using a spontaneous wireless network paradigm, it becomes apparent that this architecture is naturally more resilient in face of disasters, and is less prone to the impacts described in section \ref{sec:dependencyonfixedinfrastructure}. By running on battery power, nodes are not affected by power black-outs and damaged power cables. By using radio links the communication between devices does not suffer from broken wiring. Furthermore, by leveraging its dense deployment, and its ability to spontaneously self-organize wireless multi hop communication, the IoT brings a huge additional advantage: it comes with built-in redundancy. This means that even with a large loss of nodes, there is a good chance that the network will still consist in a giant component of physically connected nodes, which could be put to use immediately after the disaster happens.

It is however projected that IoT devices will be very diverse with respect to characteristics including computation power, memory capacity and communication capabilities. While today's cell-phones are able to transmit and receive data using Wifi, Bluetooth, UMTS or LTE with throughputs ranging from a few Mbit/s to a few hundred Mbit/s, typical wireless sensor networks (WSNs) use radio standards that provide significantly lower throughput, in the range of a few hundred kbit/s \cite{onwuka2013}. In order to ensure connectivity over large areas, it is safe to assume that any IoT device that has survived the disaster in the area may be used as potential relay. Since the available throughput is smaller than the bottleneck on the path, it may thus be that a particularly constrained IoT device severely limits the available throughput towards a given destination. Furthermore, the routing mechanisms at work in large scale spontaneous wireless network may limit this throughput even more \cite{gupta2000}.  However, in any disaster scenario, a good rule of thumb is: limited connectivity is better than no connectivity at all. When looking at the communication requirements listed in section \ref{sec:disasterrequirements}, it becomes apparent that even a low throughput, text-based emergency service would help improving the coordination, speed and efficiency of disaster response, and that the availability of such a service may save lives as a direct consequence. Such mechanisms could enable diverse services including (i) emergency broadcast to all devices in an area to warn the general public, (ii) first-responder text-based situation reports communication to central coordination instances which can then make faster and more informed decisions, or (iii) individuals may emit emergency messages which allow response forces to detect and locate them in scenarios where some people are buried for instance.

It is furthermore noteworthy that a substantial part of the IoT is expected to consist in sensors that monitor various environmental parameters, thus providing quasi-ubiquitous sensing capabilities. Using these capabilities, coupled with the resilience of the IoT may provide crucial real-time data about disaster areas, which can help decision makers to better understand the impact of a disaster and react more appropriately. Available sensor data may range from temperature readings during bush fires, radiation readings after nuclear accidents or even destruction estimates based on the number and location of nodes that become unreachable.

\section{Open Challenges}
\label{sec:challenges}

The IoT has considerable potential to contribute significantly to disaster resilience of communication networks as we discussed in Section \ref{sec:iotpotential}.
However, prior to succeeding in the `grand challenges', the IoT is challenged by a variety of open questions and unsolved problems.
Most challenges do not arise from the lack of existing technologies, but rather from a premature development of existing technologies and in particular from a lack of common standards and deployments that seamlessly interconnect. In the following section we will point out the areas where the most pressing issues arise.

\subsection{Physical connectivity and hardware limitations}
\label{sec:macphy}

Physical connectivity on a hardware level is the essential foundation to enable communication between devices. Sharing the same PHY and link layer is a requirement for data exchange between neighboring devices. For the IoT this means the use of common interface cards that  use the same radio frequencies, modulations, link layer technology etc. Multiply connected gateways are required for transitioning network technologies.

A large heterogeneity of network access technologies, though, not only increases complexity of inter-networking, but may also lead to severe deployment problems in the wireless domain. Various radios that consume interfering frequencies of the limited spectrum by incompatible technologies may harm communication capacities at large without an ability to mutually coordinate.

Mobile phones broadly use 3GPP standards for data communication, such as UMTS, and increasingly LTE. In addition, modern phones and other handheld devices (e.g., tablets) have further network interfaces such as  IEEE 802.11 (Wifi) and  IEEE 802.15.1 (Bluetooth). They are thus widespread candidates for bridging between radio technologies and serving as gateways. Similarly,  millions of Wifi access points are deployed, each of which typically featuring a wireless and a wired network interface card for offering transit from small wireless 'cells' to the remaining Internet. Energy constraints typically restrict wireless sensors to a single wireless interface, either using a link layer based on IEEE 802.15.4 or Bluetooth Low Energy (BLE), which is not backward compatible. Other gateways like IEEE 802.15.4 border routers or Bluetooth 4.0 dual-mode devices need to be in place to integrate IoT devices.

In disaster scenarios, all available devices should form a single, largely connected network---as redundant as possible. Assuming the infrastructure is down (e.g., because of power blackout or cable damage), some battery-powered devices with multiple interfaces using different radio technologies such as smartphones, tablets, laptops will have to play the role of border routers to enable physical connectivity. However, it is noteworthy that these consumer devices  typically neither have a IEEE 802.15.4 nor a BLE interface, which may lead to network partitioning because sensor networks using this link layer technology are unable to interconnect at the physical layer. 

Moreover, since 30 years the industry has focused quasi exclusively on improving infrastructure-based wireless link layer technologies. It has largely ignored spontaneous wireless networking to the point that even today---15 years after the initial 802.11 standards were published---standard Wifi ad hoc mode is often not interoperable among vendors, if  implemented at all. More generally, it remains to be seen how far new technologies can improve the performance of ad hoc, spontaneous wireless communication.

\subsection{Logical network connectivity}
\label{sec:networklayer}

The aim and outstanding success of today's Internet builds on its efficient and seamless way of interconnecting networks that use heterogeneous link layer technologies. This was achieved at large scale by using  IP (the Internet Protocol) as the unique networking protocol and TCP/UDP at the transport layer.  Wireless sensors and other constrained wireless devices are however too often based on proprietary network stacks (e.g., Zigbee or Nordic's Speedburst) that cannot interoperate across link layers or network borders. These confined networking solutions typically rely on specialized gateways to connect devices with the IP-based networks (i.e., the Internet).

Recently, the situation has improved, though, as the IETF has published relevant standards for the IoT. 6LoWPAN defines a lightweight network sublayer that enables constrained nodes (e.g., wireless sensors using IEEE 802.15.4) to interoperate natively with IPv6. 6LoWPAN thus enables a substantial fraction of IoT devices to connect directly to the Internet. It is projected that in the near future, proprietary network stacks will be phased out in favor of an IPv6 network stack using 6LoWPAN, as this brings not only benefits for vendors through standardization but also through faster time-to-market, cheaper development cycles leveraging well-known development practices and tools.

However, 6LoWPAN as a minimal standard of speaking IPv6 among devices is insufficient to orchestrate large scale spontaneous wireless networking, as required for disaster resilience described in Section \ref{sec:iotpotential}.  Improved disaster resilience relies on the ability of IoT devices to (i) dynamically reconfigure forwarding tables in order to route data over multiple wireless hops, towards destination, and (ii) dynamically adapt transport layer mechanisms to the particular versatility of multi-hop wireless communication. Both (i) and (ii) should be achieved automatically, without explicit configuration, without the need of intervention from users and network administrators, and without the help of infrastructure. Over the last decade, a significant amount of work has been accomplished in this field, which resulted in the publications of new routing protocol standards (e.g., RPL, OLSR) to cope with (i). More work is however needed to achieve better scalability of routing protocol overhead in practice---we are still far from the theoretical bounds. Furthermore, TCP modifications are desirable to efficiently accommodate multi-hop wireless communication to cope with (ii).

Additional auto configuration mechanisms are needed for IoT devices to be useful in case of disaster which results in unavailability of infrastructure-based networks. For instance, sensor networks and other IoT networks are mostly envisioned as stub networks which connect to the Internet through a given gateway. This gateway directly or indirectly determines the configuration of attached nodes, including parameters such as IP address, encryption details. Unless nodes reconfigure automatically these parameters upon detection of infrastructure unreachability, nodes that were in separate stub networks prior to the disaster may not be able to communicate with one another because the network layer will prohibit it---thus annihilating the chances of spontaneously interconnecting to form a single, large network spanning the disaster area. To efficiently enable this behavior, future work has to be carried out.

\subsection{Prioritization of data traffic}

Largely heterogeneous link transitions bear the problem of exhausting congestions that are likely to kill data flows.
Assuming the connectivity gap is bridged at the MAC/PHY layer and at the network layer as described in Sections \ref{sec:macphy} and \ref{sec:networklayer}, throughput may be very limited. The general idea is to use the available throughput for the most important services, as described in Section \ref{sec:disasterrequirements}. A challenge that remains is thus the design of mechanisms that guarantees that only these services do use of the available throughput.

An idea could be to introduce a 'disaster mode' for IoT devices. Besides their normal mode of operation, IoT nodes could switch to an alternative mode of operation in which the goal becomes spontaneous maximization of connectivity in the sense described in Section \ref{sec:iotpotential}. Furthermore, this special mode of operation could implement prioritization policies that would guarantee first responders or official organizations privileged access to the newly spawned communication network. This 'disaster mode' would be roughly comparable to the emergency call mode in today's mobile phones, where 911 calls are possible even if no registered SIM-card is activated.

In fact, such a mode of operation may be necessary anyway in case of disaster because, should massively deployed sensors and smart object resume their 'normal operation' automatically after the disaster, the limited throughput left available may be involuntarily clogged by 'unimportant' data traffic -- a case that should be avoided. Note that this may also apply to other types of data traffic, e.g., system updates on smart-phones.

As promising as such an approach sounds, there are however important additional technical questions, as well as political questions, which have to be investigated. How/when exactly would such a 'disaster mode' be triggered? What kind of regulations are needed to force vendors to integrate this mode into their devices? How should such a 'disaster mode' be standardized?

\subsection{Social acceptance}
As described in Section \ref{sec:iotpotential}, leveraging IoT devices to mitigate the impact of a disaster on network connectivity implies that devices may be required to be operated outside their intended scope, and connect to external parties that normally do not have access to those devices. For example, if privately owned sensor networks were required to relay communication traffic on behalf of governmental agencies, or on behalf of other private individuals that must send/receive emergency information, the owners of such networks would need to allow a mode of operation that they do not fully control. Social acceptance of this category of usage should be studied, to prevent situations where owners of devices actively try to block any use outside their full control---preventing in effect the approach towards more resilience.

\subsection{Network security aspects}
The IoT in general presents a number of challenges in terms of application layer and network layer security. These security challenges naturally transfer to IoT use in case of disaster scenarios. In this context, one should avoid the usual reflex of initially leaving security aspects out of the picture because ''every bit of the scarce throughput should be used for communication traffic''. 
For example, there are a number of scenarios in which unprotected network traffic could be used by malicious third parties to intentionally interrupt or alter information that is exchanged between first responders or coming from emergency calls, e.g., large-scale terrorist attacks such as 9/11. As data is routed through the IoT, attackers could try to tamper with communications ways that cripple helper organization. Furthermore, the mechanisms that trigger devices to switch to 'disaster mode' operation should itself be secure in order to prevent attacks aiming to disrupt normal network operation. These challenges are directly related to lightweight, decentralized authentication schemes.

%When looking at the above proposed emergency mode for the IoT, a vital challenge is to design a secure activation mechanism so that only legitimate authorities can actually enable this mode. This security challenge is not only of technical nature but also covers the security on organizational levels of public authorities and responsible persons.

\subsection{Towards disaster resilience}
With the technology available today, the Internet of Things cannot yet be used to improve our communication networks resilience in face of large-scale disasters. Several challenges must be addressed beforehand. While from a technical perspective the open questions we have identified yield substantial issues to be solved there are no fundamental show-stopper to allow the IoT to mitigate the impact of a disaster on network connectivity. The main question is thus not whether the IoT can be leveraged to improve disaster resilience, but rather to which extent and how it should be adopted.

\section{Conclusions}
\label{sec:conclusions}

The Internet of Things is already here. Beyond traditional routers and Internet hosts such as PCs or smartphones/tablets, a new category of battery-powered, connected machines has emerged, and applications using these machines are announced and brought to the market on a daily basis. Projections indicate that massive deployment of such devices is dawning, and will soon revolutionize the edge architecture of the Internet, by leveraging not only infrastructure-based wireless networking but also spontaneous wireless networking. This enriched architecture can significantly improve the resilience of basic data communication services in face of disasters that damage conventional communication network infrastructure. 

While the IoT is not able to provide the full range of communication services expected from pre-disaster Internet, one can nevertheless envision providing better-than-nothing services such as emergency micro-messaging, using IoT devices as relays and popular handheld devices (e.g. smartphones) as user terminals. This paper proposed an overview of this vision, and highlighted the major advantage such an approach could bring: the automatic reconfiguration of the network to interconnect surviving devices immediately after the disaster, even if the infrastructure is down and the power grid is out. Basic connectivity and simple text-based data communication could then remain available during the crucial gap between the time when the disaster occurs and the time when qualified manpower reach the area and set up dedicated hardware putting conventional communication infrastructure back in service. 

There are however a number of challenges that need to be addressed before this vision can be realized. This paper provided an analysis of the different categories of issues that lie ahead. These concern on one hand technical aspects such as physical connectivity requirements, network protocol stack enhancements, or data traffic prioritization schemes, and on the other hand non-technical aspects such as social and political considerations. We argue that while the relevant technical issues are substantial, there are no identified show-stoppers. Concerning non-technical aspects, we argue that legislating on the matter would probably be necessary. We propose the definition of a mandatory 'disaster mode' of operation for IoT devices (similar to cellphone's 911 mode of operation), which could automatically kick in to reconfigure the surviving network elements in cases where infrastructure is out of service, enabling automatically basic connectivity and simple text-based data communication for emergency purposes.

\ \\ \ \\

\begin{small}
\bibliographystyle{IEEEtran}
\bibliography{references}
\end{small}

\end{document}